\documentclass[a4paper]{spie}  
\usepackage[]{graphicx}

\title{A multiplexer for the ac/dc characterization of TES based
  bolometers and microcalorimeters.} 

\author{L. Gottardi\supit{a}, H. Akamatsu\supit{a}, M. Bruijn\supit{a},
  J.R. Gao\supit{a}\supit{b}, R. den Hartog\supit{a},
  R. Hijmering\supit{a}, H. Hoevers\supit{a}, P. Khosropanah\supit{a},
  J. van der Kuur\supit{a}, T. van der Linden\supit{a}, M. Lindeman\supit{a}, M. Ridder\supit{a} 
\skiplinehalf
\supit{a} SRON National Institute for Space Research, \\ Sorbonnelaan 2, 3584 CA Utrecht, The Netherlands \\
\supit{b} Kavli Institute of NanoScience, Faculty of Applied Sciences,
 Delft University of Technology,
Lorentzweg 1, 2628 CJ Delft, The Netherlands
}

\begin{document} 
  \maketitle 

\begin{abstract}
At SRON we are developing the Frequency Domain Multiplexing (FDM) for the
read-out of the TES-based detector array for the future infrared and X-ray space
 mission.
We describe the performances of a multiplexer designed to increase the experimental throughput in the characterisation of ultra-low noise equivalent power (NEP)
 TES
bolometers and high energy resolving power X-ray microcalorimeters arrays under
ac and dc bias. We discuss the results obtained using the TiAu TES
bolometers array fabricated at SRON with measured dark NEP below  $5\cdot 10^{-19} \mathrm{W/\sqrt{Hz}}$ and saturation power of several fW. 

\end{abstract}


\keywords{FDM. cryogenic detector,TES, infrared bolometers, SQUID, x-ray microcalorimeter}

\section{INTRODUCTION}
\label{sec:intro}  

We developed a Frequency Domain Multiplexer (FDM) in order to increase the
experimental throughput in the characterisation of TES-based X-ray
microcalorimeters and bolometer array under ac bias at frequency
higher than 1 MHz.  To simultaneously measure a large number of pixels
a baseband feedback scheme \cite{Hartog12} is required. However, to
perform a single pixel characterisation, each ac channel can be
read-out sequentially in time. In this way the SQUID amplifier dynamic
range is less critical and the read-out of a single low-G bolometers,
generally  operating at TES rms current values lower than $2\mu
\mathrm{A}$, can be done in open loop. The X-ray TES microcalorimeters
operate at  larger current then the bolometers. It that case baseband feedback will be used. 

Another reason to develop a new FDM set-up was to understand the
energy resolution degradation observed with X-ray microcalorimeters
under AC bias \cite{Gottardi12xray} 
So far the ac bias experiments where carried on at
frequencies below 700kHz, using sub-optimal
circuit elements such as high-inductance SQUID amplifiers with strong
coupling between feedback and input coil, low-Q
discrete LC resonators and large stray inductance and resistance
connections. It has been shown that operating the TES under ac bias at
too low a frequency can lead to a degradation of the energy resolution
\cite{JvdK02}. Only recently high-Q lithographic, superconducting LC
filters\cite{Bruijn12}, tuned to frequencies between
2 to 5MHz, and high dynamic range, low input inductance SQUID became
available for the FDM read-out at MHz. This novel technology has been
implemented in the FDM set-up. 

Moreover, the experiments set-up used so far suffered from a
poor magnetic shielding. We greatly improve the
magnetic field shielding of our single pixel test bed.  The optimal
shielding design and the accurate
selection of non-magnetic components guarantees an uniform and low ($<
1\mu$Tesla) magnetic field environment at the detector stage.   

With the new FDM set up we will be able to characterise TES based detectors
under  optimised experimental condition using circuit component tailored for FDM.
 Here below we present the results of the single pixel characterisation under ac bias of a low-G bolometer developed for Safari \cite{Pourya12,Hijmering12}.  


\section{EXPERIMENTAL SET-UP} 
\label{sec:setup}

The multiplexers described below was developed to be used both with
ultra-low noise equivalent power (NEP) TES bolometers  and with high energy
resolving power x-ray microcalorimeters. The former requires very low
parasitic power loading ($<1\mathrm{fW}$), which is achieved by means of light blocking
filters in the signal loom feedthrought and a  light-tight
assembly. The x-ray microcalorimeters are very sensitive to magnetic
fields and their performance is optimal at static field lower than  
$1\mu \mathrm{Tesla}$).
 Special care has been taken to design the magnetic
shielding and to improve the uniformity of  the applied magnetic field
across the array. 
The FDM multiplexer consists of a low magnetic impurity copper bracket
mounting on top  the SQUID array, the LC
filters and the TES array chips and a printed circuit board for the electrical
connection. One thermometer and
one heater are glued at the bottom of the copper plate for the read-out and stabilisation of the temperature. 

\begin{figure}[htbp]
    \centering
    \includegraphics[height=6cm]{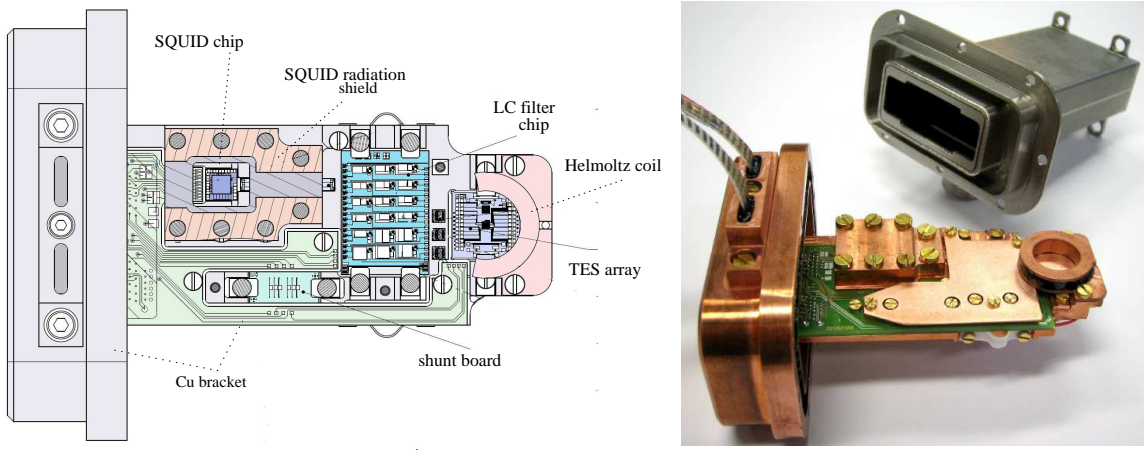}
    \caption[example] 
   { \label{fig:FDMscheme} A schematic drawing and a picture of the
     FDM set-up.}
   \end{figure}

The TES arrays chip fits into a superconducting Helmholtz coil fixed
at one end of the bracket. The coil is used to generate an
uniform perpendicular magnetic field over the whole pixels array. 
A schematic drawing and a picture of the set up is shown in Fig.~\ref{fig:FDMscheme}.
The shielding of the external magnetic field was achieved by fitting the
bracket into a Nb can wrapped by few layers of metallic glass  tape. The matching of the Nb can with the bracket lid was designed such
that it forms a labyrinth, filled with carbon loaded
epoxy on the copper lid side. In this configuration the Nb can provides both the
required magnetic and stray light shielding.

The electrical connections from the cold stage of the cooler to the
SQUID array and TES bias circuit elements are achieved by means of
two superconducting looms fed through a narrow 10 mm long channel filled with carbon loaded epoxy.
 The circuitry PC-board currently allows to read-out  18 pixels under ac 
 bias in a FDM configuration and 2 pixels under dc or ac bias with the
 SQUID amplifier located outside the Nb can.   

For the FDM read-out we use a low noise two-stage  PTB SQUID array current sensor with
on chip linearization \cite{Kiviranta08}, low input inductance
($\mathrm{L<3 nH}$) and low power dissipation ($\mathrm{P<20 nW}$) \cite{Drung09}. 
The SQUID amplifier chip  is placed in a
radiation shielding cavity whose inner side is coated with a 2 mm
thick radiation absorber made from carbon loaded epoxy with mixed  SiC grains
with  size ranging  from $100 \mathrm{\mu m}$  to 1 mm \cite{epoxy1,epoxy2}. This
precaution was taken to minimise possible  loading of the bolometers
due to Josephson radiation, typically in the range of 4-8 GHz, emitted
by the SQUID junctions. 
The SQUID chip is thermally coupled to the
bracket by means of several Au bonding. The
electrical connection from the SQUID chip to the LC filters is done by
means of Nb strip lines on a 20 mm long interconnection chip. These lines
act as  a low-pass filter with a calculated roll-off around 500 MHz. 
We operate the SQUID in open loop. The output signal is
  amplified by a 20 MHz bandwidth, low input voltage noise,  commercial
  electronics (Magnicon GmbH). 

The TES array chip is connected via a superconducting interconnection
chip to the lithografic high-Q
 LC resonators arrays developed at SRON \cite{Bruijn12}. The nominal inductance of
 the coil used in  each filter is $L=1 \mu\mathrm{H}$, while the
 capacitances C are designed such that the  frequencies
 $f_0=\frac{1}{2\pi\sqrt{LC}}$ are spread at a constant interval of
 about 200kHz in the range from 2 to 5 MHz. The LC filter chip used had 18
 filters of which 14 were properly functioning.

\section{FDM READ-OUT CHARACTERISATION} \label{sec:fdmreadout}

We evaluated the performance of the FDM read-out by measuring the
SQUID input current noise and dynamic range, the quality
factor of the LC filters and the uniformity and goodness of the
magnetic shielding.  

\subsection{SQUID amplifier and LC filters test}

\begin{figure}[htbp]
\centering
\includegraphics[width=1.0\textwidth,keepaspectratio, angle=0]{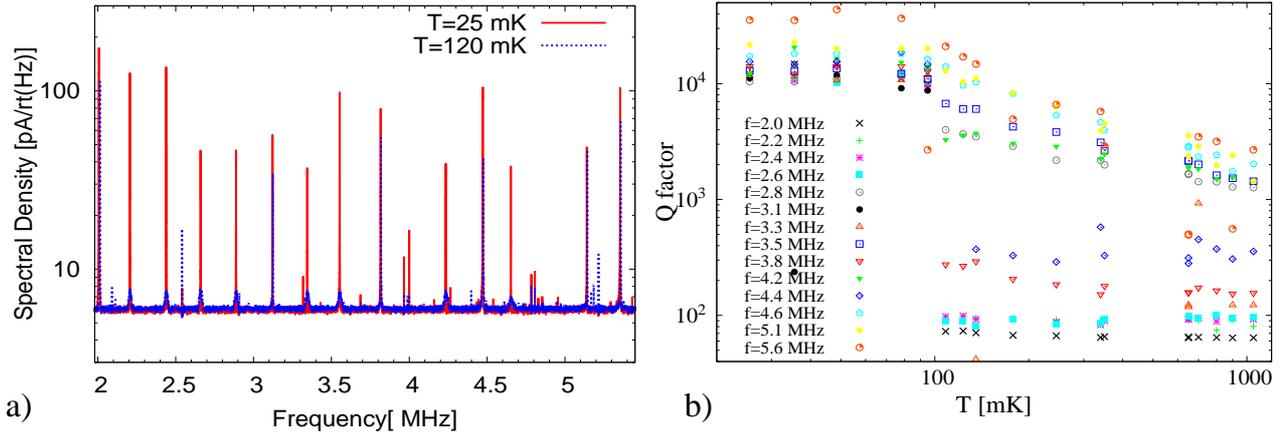}
\caption{ {\bf a)} Squid current noise with 14 high-Q lithographic
  resonators connected. We connected 8 low-G bolometers, with a normal resistance
$R_N=160 \mathrm{m}\Omega$ and $T_C=100 \mathrm{mK}$ in
  series with  the LC filters
with resonance frequencies of 2.2, 2.4, 2.6, 2.8, 3.4, 3.6, 4.2 and
4.6 MHz respectively. The other resonators were shorted using
superconducitng Al bonding. {\bf b)} The resonances Q-factor as a function of temperature. \label{SQUIDnoise}}
\end{figure}

In Fig.~\ref{SQUIDnoise} we show the SQUID input current spectral density 
during nominal operation of the multiplexer. 

For the measurements described here we connected 8 low-G bolometers, with a normal resistance
$R_N=160 \mathrm{m}\Omega$ and $T_C=100 \mathrm{mK}$, to 8 LC filters
with resonance frequencies of 2.2, 2.4, 2.6, 2.8, 3.4, 3.6, 4.2 and
4.6 MHz respectively. The other filters were
superconducting shorted using Al bonding. 
At temperature $T=30 \mathrm{mK}$ we
  measured a SQUID input current noise of $6
  \mathrm{pA/\sqrt{Hz}}$ (with $1/M_{in}=19.6\mu\mathrm{A/\Phi_0}$) over the whole interesting
  frequency range from 2 to 5 MHz, as expected for this particular
  SQUID chip. All the
  resonators had a Q factor larger than $10^4$ at $T<100 \mathrm{mK}$.
 The SQUID operates in a linear regime for peak-peak input currents lower than
 $12 \mu \mathrm{A}$ corresponding to a dynamic range of $I_{tes,pkpk}/I_{ns}=2\cdot10^6\sqrt{Hz}$ . 

\subsection{Magnetic shielding}

\begin{figure}[htbp]
\centering
\includegraphics[width=1\textwidth,keepaspectratio, angle=0]{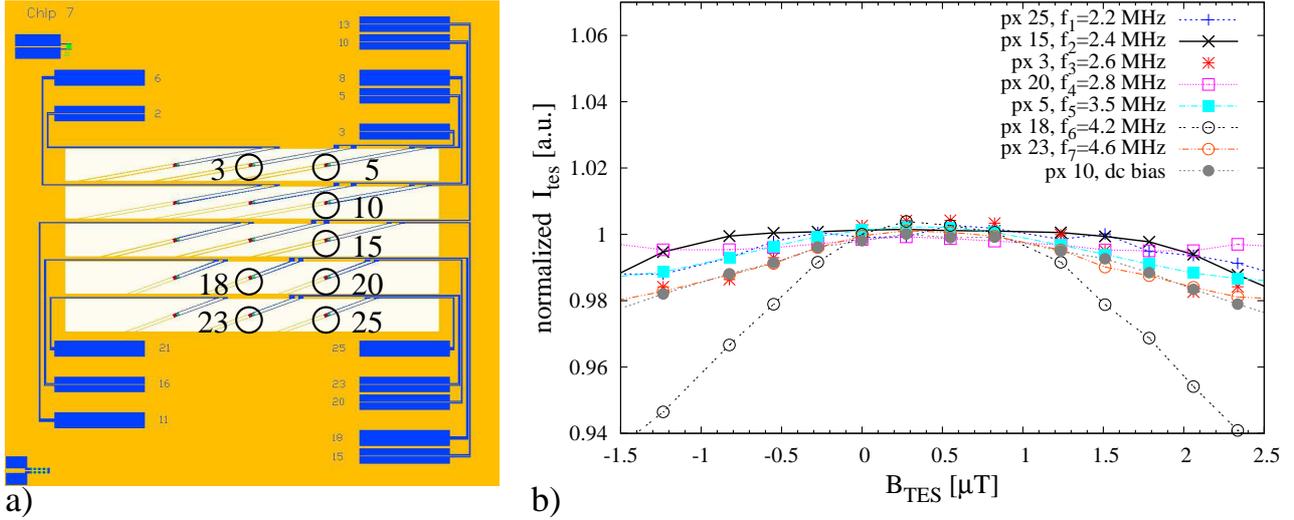}
\caption{ {\bf a)} Mask layout of the array under test showing the
  pixels location. {\bf b)} TES current response of 8 pixels in the array  as a function
  of applied magnetic field. As the maximum is found at the same
  magnetic field value  $B_{res}=+5\cdot 10^{-7}\mathrm{T}$, for each
  pixel, we conclude that the B-field shielding and the B-field bias
  is uniform within $\pm 1.5\cdot 10^{-7}\mathrm{T}$.  \label{IpxlsvsB}}
\end{figure}

In order to test the quality of the magnetic shielding we measured the 
current response of several TES pixels, distributed over a surface of $3.3\times 3.3 \mathrm{mm}$ in diameter,  as a function of the magnetic
field $B_{TES}$ applied perpendicularly to the TES. The results are
shown in Fig.~\ref{IpxlsvsB}. For each measured
pixel the response is symmetric around the applied magnetic
field $B_{res}=(4\pm 1.5)\cdot 10^{-7}\mathrm{T}$. 
This value is an estimation of the residual magnetic field
perpendicular to the TES. 

From the graphs we estimate that the B-field shielding and the B-field bias is uniform within $\pm 1.5\cdot 10^{-7}\mathrm{T}$, 
In more than 6 cryogenic runs the measured residual
magnetic field has always been better than $B_{res}=|1\cdot 10^{-6}| \mathrm{T}$.




\section{SINGEL PIXEL CHARACTERIZATION WITH THE FDM SET-UP} \label{sec:singlepxlac}

In the following session we present the results of a single pixel 
characterisation using the readout channel at 2.4MHz.

\begin{figure}[h]
   \begin{center}
   \begin{tabular}{c}
   \includegraphics[height=4.4cm]{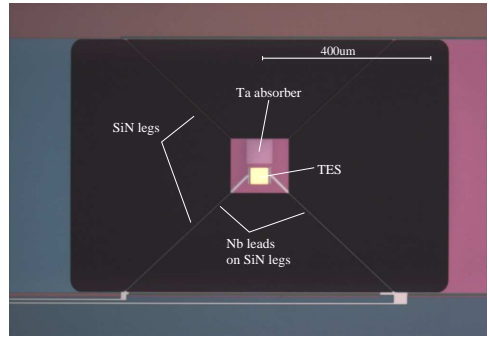}
   \end{tabular}
   \end{center}
   \caption[example] 
   { \label{fig:fotopixel} A picture of the low-G TES bolometer used
     for the single pixel characterisation under ac bias .}
   \end{figure}

The device under test is a low-G bolometer based on a Ti/Au (16/60 nm)
bilayer, deposited on $0.5 \mu \mathrm{m}$ thick suspended SiN membrane.
The TES area is $50\times 50 \mu \mathrm{m^2}$.  It has a critical
temperature of $T_C=78.5\,\mathrm{mK}$, a  normal state  resistance  of
$R_N=98\,\mathrm{m\Omega}$, a measured $G=0.27$pW/K and a calculated NEP of $2.3\cdot 10^{-19} \mathrm{W/\sqrt{Hz}}$. 
An 8 nm thick Ta absorber with an area of $70\times70 \mu \mathrm{m}^2$ is deposited close to the TES. The absorber and
the TES are sitting on a $130\times 70 \mu \mathrm{m}^2$ SiN island.
Fig.~\ref{fig:fotopixel} shows a picture of the device. 

There are 4 SiN cross-shaped
supporting legs that are $2 \mu \mathrm{m}$ wide and $400 \mu \mathrm{m}$ long.
The electrical contact to the bolometer is realized by 90 nm thick
Nb wiring on the top of SiN legs. 
The sensor was previously characterised
under dc bias in the  set-up described in \cite{Pourya12,Hijmering12}  and showed a power plateau of
$9.4\mathrm{fW}$ and a dark NEP of $4.8\cdot 10^{-19} \mathrm{W/\sqrt{Hz}}$ at 30 mK.
The pixel is fabricated on a test chip with 4 other pixels.

\subsection{Current-voltage and power-voltage characteristics}

\begin{figure}[h]                                                              
 \center
 \includegraphics[width=1.\textwidth,angle=0]{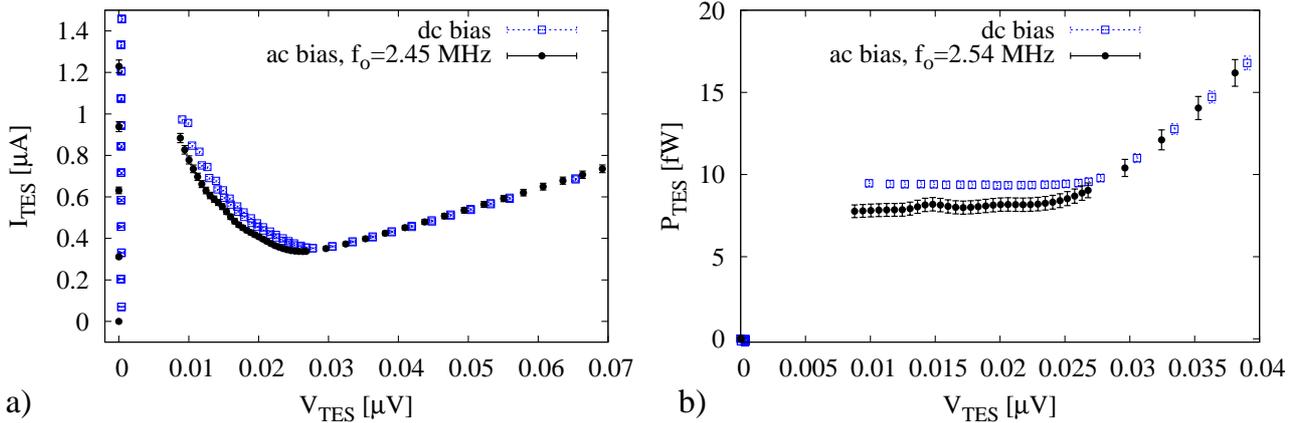}\caption{TES
   current-to-voltage (I-V) and
power-to-voltage (P-V) characteristics for pixel C24-P05 measured at a bath
temperature of T=30 mK. \label{IVPVacdc}}
\end{figure}

In Fig.~\ref{IVPVacdc} we show the TES current-to-voltage and
power-to-voltage characteristics measured both under ac and dc bias  at a bath
temperature of T=30 mK and at zero perpendicular magnetic field. 

Under ac bias the  saturation power  is
$7.2 \mathrm{fW}$, which is about 2 fW lower than the power observed in
the dc bias measurement set-up. 

The SQUID amplifier mounted inside the light-tight box in the FDM
set-up is a possible source of power loading. The power dissipated in
the SQUID at its optimal working point is about 15 nW. In order to
quantify the amount of power leaking from the SQUID to the bolometer
array we measured the power-voltage characteristic of another
pixel in the array using an auxiliary ac read-out system. 
In this case the TES signal is measured by a SQUID array hosted in a module
mounted externally to the light-tight box. 
When switching the FDM SQUID array on and off we observed a difference
of about 0.5 fW in the pixel saturation power, which is not sufficient
to fully explain the parasitic loading at 30mK. 


At bath temperatures $T>70 \mathrm{mK}$ the saturation power measured
under ac and dc bias was the same within the measurement
uncertainties. The latter observation excludes the presence of an electrical loading since the read-out conditions are
identical at all bath temperatures.
A possible explanation for the lower power plateaux
measured at $T=30 \mathrm{mK}$ in the FDM set-up could be   a bad
thermalization of its circuit components. A difference of about 15 mK
between the measured temperature of the copper bracket and the real
temperature of the bolometer array Si chip is sufficient to
explain the discrepancies observed.

We believe that this issue can be solved with minor improvement of the
experimental setup and does not put any fundamental constrain in
the performance of a low-G bolometer under ac bias.

\subsection{Noise measurements}

\begin{figure}
   \begin{center}
   \begin{tabular}{c}
   \includegraphics[height=6cm]{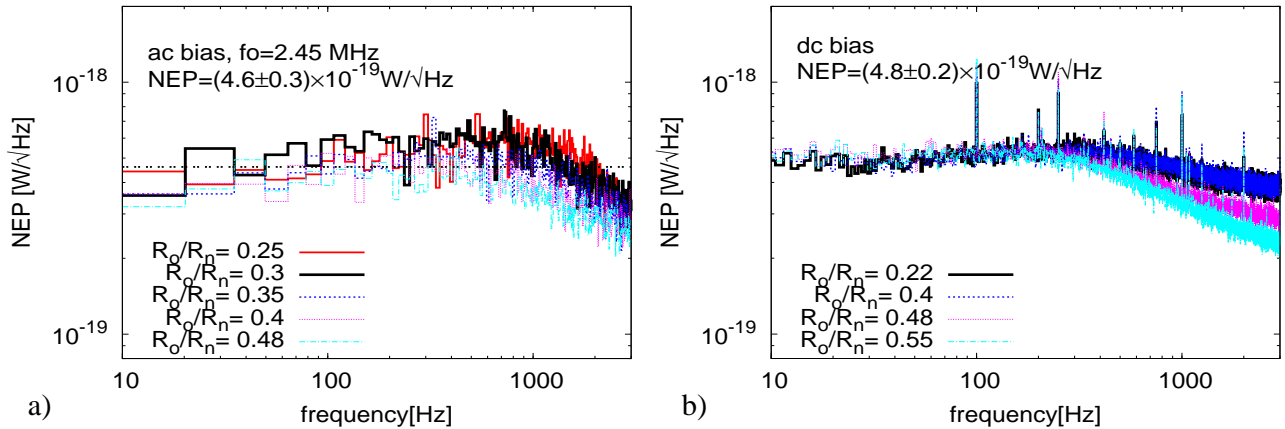}
   \end{tabular}
   \end{center}
   \caption[example] 
   { \label{fig:nepACDC} 
Measured dark NEP al low frequency for different bias point
respectively with the ac ({\bf a)}) and dc ({\bf b)}) bias read-out.}
   \end{figure}

Fig.~\ref{fig:nepACDC} shows the dark NEP  spectra at several
bias points in the transition for a bath temperature of 30 mK. 

Both the results obtained under ac and dc bias are shown. 
The dark NEP was calculated by dividing the
current noise by the responsivity at low frequency. which can be
approximated by $\frac{1}{I_O(R_O-Z_{th})}$, where $I_O$ is the
effective bias current, $R_O$ is the TES resistance and $Z_{th}$ is the Thevenin
impedance in the bias circuit as derived from the calibration of the
I-V curves.
The measured dark NEP at low frequency at all working points in the
transition was about $(4.6\pm 0.3)\cdot 10^{-19} \mathrm{W/\sqrt{Hz}}$ and
$(4.8 \pm 0.2)\cdot 10^{-19} \mathrm{W/\sqrt{Hz}}$ respectively for
the ac and dc bias case.


\section{CONCLUSION} 

We fabricated and tested an FDM set-up for the single pixel
characterisation of TES based bolometer and microcalorimeters under ac
bias in the  MHz frequency range. The set-up is designed to read out
18 pixels under ac bias and 2 pixels under ac or dc bias.

We validate the performance of the ac bias read-out  by characterising the
SQUID amplifier, the lithographic high-Q LC filters and the magnetic
 shielding at $T=30\mathrm{mK}$. The SQUID
 amplifier has an input current noise of $6 \mathrm{pA/\sqrt{Hz}}$ and a dynamic range of
  $1.7\cdot 10^6\sqrt{Hz}$ over the whole interesting
  frequency range from 2 to 5 MHz. 
  All the measured superconducting LC resonators has a Q-factor larger than $10^4$ at $T<100 \mathrm{mK}$.
  We measured a  static magnetic field of the order of $5\cdot
  10^{-7}\mathrm{T}$ inside the Nb shield at the array location and
  perpendicular to the pixels. The residual field is uniform within $\pm 1.5\cdot 10^{-7}\mathrm{T}$  over an area of $3.3\times 3.3 \mathrm{mm}$ and found to be
reproducible in more than six cryogenic runs. 

We fully characterised a low-G device at an ac bias frequency of 2.5
MHz. The TES pixel was previously measured under dc bias in a
different cryogenic set-up.
The pixel showed comparable noise feature when operating under ac and
dc bias. A dark NEP of $(4.6\pm 0.3) \cdot 10^{-19}
\mathrm{W/\sqrt{Hz}}$ was observed at 30 mK with the ac bias
readout. The detected saturation power was $7.2\mathrm{fW}$ under ac
bias, which was about 2 fW lower than the power observed in the dc bias measurement set-up.
We believe this was due to a bad thermalization of the LC filter and
bolometer array chips and that this issue can be solved with minor
improvement of the experimental setup. From the measurements reported
here we therefore conclude that the behaviour of the low-G bolometers
operating under dc bias and ac bias at a frequency of 2.4MHz is equal.  

A similar FDM set-up is currently being assembled to perform single
pixel characterisation of TES-based X-ray microcalorimeter array.


\acknowledgments     
 
We thank Marcel van Litsenburg, Lou Verhagen and Robert Huiting for their precious technical help.


\bibliography{GottardiSPIE2012}   
\bibliographystyle{spiebib}   

\end{document}